\documentstyle[epsf,12pt]{article}
\begin{document}
\renewcommand{\baselinestretch}{1.5}

\newcommand\beq{\begin{equation}}
\newcommand\eeq{\end{equation}}
\newcommand\bea{\begin{eqnarray}}
\newcommand\eea{\end{eqnarray}}

\newcommand\rRo{\rho_{R0}^A}
\newcommand\rLo{\rho_{L0}^A}
\newcommand\rR{\rho_{R}^A}
\newcommand\rL{\rho_{L}^A}

\newcommand\al{\alpha}
\newcommand\sig{\sigma}
\newcommand {\dlt}{ \frac{\delta}{\pi}}
\newcommand {\dlts}{\frac{\delta^2}{\pi^2}}

\newcommand\pisig{\Pi_{\sigma i }}
\newcommand\pisigm{\Pi_{I\sigma i }}
\newcommand\sumsig{\sum_{\sigma i }}
\newcommand\psig{p_{\sigma i }}
\newcommand\xsig{x_{\sigma i }}
\newcommand\xmsig{x_{-\sigma j }}
\newcommand\sumI{\sum_{I}^{N}}

\newcommand\sumi{\sum_i}
\newcommand\sumj{\sum_j}
\newcommand\sumJ{\sum_J}

\newcommand\expo{e^{iS(\{x_{+i}\}, \{x_{-j}\})}}
\newcommand\expon{e^{-iS(\{x_{+i}\}, \{x_{-j}\})}}

\newcommand\tp{\tilde\phi}
\newcommand\dpi{\delta/\pi}
\newcommand\ddpi{\delta^2/\pi^2}
\newcommand\px{\partial_x}
\newcommand\pt{\partial_t}
\newcommand\prl{Phys. Rev. Lett. }
\newcommand\prb{Phys. Rev. {\bf B}}

\newcommand \Dlt {\bar\delta}
\newcommand \OO {\tilde O}
\newcommand \dpp {\partial_x\phi_+ }
\newcommand \dpm {\partial_x\phi_- }
\newcommand \dppi {\partial_x\phi_{+i} }
\newcommand \dpmi {\partial_x\phi_-{i} }

\newcommand \dpc {\partial_x\phi_c }
\newcommand \dps {\partial_x\phi_s }

\newcommand \dpmp {\partial_x\phi_{\mp} }

\newcommand \dtp {\partial_x\theta_+ }
\newcommand \dtm {\partial_x\theta_- }
\newcommand \dtpi {\partial_x\theta_{+i} }
\newcommand \dtmi {\partial_x\theta_{-i} }

\newcommand \dtc {\partial_x\theta_c }
\newcommand \dts {\partial_x\theta_s }

\newcommand \npl {n_{+L}}
\newcommand \npr {n_{+R}}
\newcommand \nml {n_{-L}}
\newcommand \nmr {n_{-R}}
\newcommand \npml {n_{\pm L}}
\newcommand \npmr {n_{\pm R}}
\newcommand \hf {\frac{1}{2}}

\hfill MRI-PHY/P991031

\centerline{\bf Transport in a class of exactly solvable models of 
interacting fermions}
\vskip 1 true cm

\centerline{P. K. Mohanty \footnote{{\it e-mail address}:  
peekay@mri.ernet.in} and Sumathi Rao \footnote{{\it e-mail
address}: sumathi@mri.ernet.in}}  
\centerline{\it Mehta Research Institute, Chhatnag Road, Jhunsi,}
\centerline{\it Allahabad 211019, India.}

\vskip 2 true cm
\noindent {\bf Abstract}
\vskip 1 true cm

We study transport in a class of exactly solvable models of interacting
fermions in one dimension. We contrast these models with models of
non-interacting fermions in an Aharanov-Bohm ring to which they are
superficially similar. We introduce magnetic and non-magnetic
impurities
at a site, through either a weak $\delta$-function potential
or through a weak link. Using a renormalisation group analysis,  
we show that the strength of the nonmagnetic impurity is not affected
by the interaction, whereas the magnetic impurity cuts the wire at the
impurity site.

\vskip 1 true cm


\newpage

Strongly correlated transport is an emerging field, which has
suddenly shot into prominence in the last few years. One reason
has been the advances in semiconductor technology, which have 
enabled the fabrication of one-dimensional quantum wires in the
single channel limit\cite{1DWIRE}. Besides,  
the recent discovery of carbon nanotubes\cite{CTUBES} and in particular, the 
actual fabrication of single-wall  carbon nanotubes and multi probe
experiments\cite{DEKKER} on them have spurred both 
theoretical and experimental
activity in this field.

One dimensional interacting electrons are known to exhibit Luttinger
liquid (LL) behaviour, rather than Fermi liquid behaviour, characterised by
spin-charge separation and  absence of Landau
quasi-particles\cite{HA}. 
In the
presence of impurities, transport responses are marked by interaction-
dependent power laws.  The dramatic effect of interactions on transport 
in a LL was shown in 
a 
paper  by Kane and Fisher\cite{KF}, who studied
transport in LLs, through both single and double barrier
structures. They showed  through renormalisation group arguments
that spinless electrons with repulsive interactions, incident upon a single
barrier are completely reflected at zero temperature and for 
vanishing bias, whereas for attractive interactions there is perfect 
transmission. This analysis has been further modified by 
several groups\cite{GR}
who included other corrections, but essentially verified their
analysis. It has also been verified by an exact solution at  
arbitrary coupling constant using the thermodynamic Bethe
ansatz\cite{LUDWIG}.
Furthermore, 
their analysis has also been extended to other cases, such
as with addition of external biases\cite{EG}, crossed LLs\cite{KE} and
double and multiply crossed  LLs\cite{DR}.

All of these analyses, are essentially within the bosonised
Luttinger model approach, with strong or weak barriers, whose effects
are then extrapolated using renormalisation group arguments. 
Since the difference between transport for the non-interacting
Fermi liquid and the interacting LL is so dramatic,
it is worthwhile to see whether such results are  
reproduced in other models of interacting fermions. 
With a view to addressing this
problem, in 
this paper, we study transport in a specific exactly solvable 
model\cite{SS} of interacting fermions in one dimension. This model is 
described in terms of two species of
fermions, with pseudospin index $\sigma = \pm$ and with 'gauge'
interactions described by a Hamiltonian given by
\beq
H =  \sumI \sumsig a_{I} (\pisig)^{2I}.
\label{one}
\eeq
Here, $\pisig = \psig + \sigma A_{\sigma}(\xsig)$ is the `covariant
momentum' introduced in Ref.\cite{SS} and  $N$ is a  `band index'
introduced in Ref.\cite{US,SS2}, which generalised these models 
by allowing for higher
powers of the covariant momenta in the Hamiltonian. 
Particles interact with each other
via the 
gauge potential, given for the particle at the position $x$ by
$A_{\sigma}(x) =\sumj V(x-\xmsig)$ - $i.e.$, the potential for a
particle with positive pseudospin  is due to the presence of the particles
with negative pseudospin and vice-versa. 
Note that the potential
depends on {\it all} the particles of the opposite sign of pseudospin
irrespective of the number of bands. The potential is chosen to be an
even function, vanishes at infinity and explicitly breaks time-reversal
invariance, although it is invariant under a combined operation of 
time reversal
and reversal of pseudo-spin index. In this paper, we shall confine
ourselves to the original model of Schulz and Shastry\cite{SS} and set
$N=1$ and $I=1$.

To study the problem of transport in this model, we first note that we
cannot study the model on an open wire as is normally done for
transport problems, because here interactions are essentially
introduced through a magnetic flux. Particles of one species 
(or pseudospin orientation)
give rise to an effective Aharanov-Bohm flux acting on the
particles of the other species(other pseudospin orientation).
Such interactions cannot
be introduced in an open wire. Hence, to study equilibrium transport in these
models, we shall consider fermions on a ring of length $L$ -$i.e.$,
we impose the periodic boundary conditions $\psi(x) = \psi(x+L)$. 
Instead of measuring current in the limit of vanishing external
voltage, we shall measure the persistent current in the model in the
limit of vanishing external flux through the ring.

By performing a pseudo-unitary transformation on the model, 
\beq
\expo p_{\sigma i } \expon = p_{\sigma i} - \partial_{x_{\sigma i}}
 S(\{x_{+i}\},\{x_{\-j}\}),
\eeq
the interaction term in Eq.(\ref{one}), is
eliminated and the Hamiltonian is transformed to the free one given by
$H=(1/2)\sumsig(\psig)^{2}$. However, the boundary
conditions on the wave-functions are now different and the
quantisation conditions for the momenta are now replaced by
\beq
L k_{\pm,i} \mp N_{\mp}\delta = 2\pi n_{\pm,i}.
\eeq
Here $\delta$ is a
phase shift that can be computed in terms of the unitary
transformation function $S$, which in turn is related to the original
interaction, since $S$ is chosen to cancel the interaction\cite{SS}.
$n_{\pm,i}$ are quantum numbers analogous to those used in the
non-interacting case.
Note that $\tilde\delta$, which  is
defined as  the fractional part of
${\delta\over 2\pi}N_{\pm }$ ($\tilde\delta$ takes value from $-.5$ to
+.5) leads to interactions;  the
integer part of the ostensible interaction term ${\delta\over 2\pi}
N_{\pm }$ merely gives integer changes in the quantum numbers
$n_{\pm,i}$,  which
anyway take values over all integers.

When there are no interactions, -  $i.e.$, when we have free electrons
on the ring(or  ${\delta\over 2\pi}N_{\pm }$ is an integer),
 - and when there is no external flux through the ring, 
clearly there is no current. 
However, once we introduce interactions, 
either by changing $\delta$ or the number of particles, 
essentially, an intrinsic  flux is introduced through the ring, which
causes a bias. 
Thus, if we plot
the persistent current as a function of the
external flux as is normally done, we find a shift in the plot with 
the maximum of the current
occurring  when $\phi_{\rm ext} =n\phi_0/2 - \tilde\delta$ 
and the minimum or zero
current occurring  at $\phi_{\rm ext}=n\phi_0 - \tilde\delta$,
where $n=$
integer and $\phi_0$ is the flux quantum. This implies a non-zero
magnetic moment of the interacting ring, even in the absence of any
external driving flux. In this model, this is not surprising, since
the model intrinsically violates the discrete
time reversal (TR) symmetry.
 
Our aim is to study transport in this theory at low temperatures and
at low external driving forces (here an external flux), when  an impurity 
is introduced in the model. Since for open wires, interactions dramatically
change the  results\cite{KF} 
we expect the differences between interacting and non-interacting fermions
on the ring also to be dramatic for transport through barriers or 
constrictions. How do we see this ?
Once any impurity is introduced
at a site, the model ceases to be exactly solvable. Since one has a first
quantised formulation of the model, one may expect that a
Landauer-type scattering matrix approach to transport (which is applicable
for free fermions) may be a viable
way to study impurity scattering, although here the fermions are
interacting. However, we find that the infrared catastrophe,
which changes the states of {\it all} the particles in the system 
whenever a single
particle is added to or removed from the system, prevents a convenient
analysis in the first quantised formulation. In other words, because of
the strong interaction, which implies a collective motion of all the
particles, it is impossible to isolate a single particle, study its
transmission and reflection, and then add up the contributions of all
the particles as is done in the Landauer-Buttiker formalism.
Hence, we follow the
more standard procedure of first bosonising\cite{BOS} the low energy
(collective) excitations, then rewriting the impurity scatterer in terms of
the bosonic
variables and studying its effect perturbatively,  and finally using 
renormalisation group to extrapolate the results.  

We first obtain the ground state energy $E_0$ in a sector
with $N_{\pm}$ particles by choosing $n_{\pm,i} = n_{\pm,i}^0 \pm
[[{\delta\over 2\pi} N_{\mp}]]_{\rm int}$, where $n_{\pm,i}^0$ are the
quantum numbers in the absence of any interaction. Excitations about
the ground state are obtained by constructing the low energy
Hamiltonian when particles are added to the left and right Fermi
points. 
Let us
assume that the ground state has $2n_0+1$ particles of each kind and
that $\frac{\delta}{2\pi}(2n_0+1)$ is an integer.
The addition
of $\npmr(\npml)$ particles at the right(left) Fermi points will cause 
a second order change in
the ground state energy given by 
\bea
E^{(2)} &=& \frac{n\pi^2}{4L}\{(1+\Dlt^2)(N_+^2 + N_-^2) + (J_+^2 + J_-^2)
+ 2\Dlt( J_+N_--J_-N_+ )\}
\eea
where the initial particle density is denoted as $n=\frac{2(2n_0+1)}{L}$,  
$\Dlt=\frac{\delta}{\pi}$, and $N_{\pm}= \npmr +\npml$ and $J_{\pm} = 
\npmr-\npml$ are the particle and current quantum numbers respectively.

Let us introduce boson fields $\phi_{\pm}$
and their conjugate 
momenta $\Pi_{\pm}= \partial_\tau\phi_\pm = \partial_x\theta_\pm $, 
following the notation of Shankar 
\cite{SHANKBOS}. These are related to the current and charge densities
as  
\beq
N_{\pm}=\frac{L}{\sqrt\pi}\partial_x\phi_{\pm} ~~~ {\rm and} ~~~
 J_{\pm}=-\frac{L}{\sqrt\pi}\partial_x\theta_{\pm} . 
\eeq 
So, the
effective low energy Hamiltonian including quantum fluctuations is now
given  by 
\beq
H =\frac{n\pi}{4}\int dx \{ (1+\Dlt^2)[ (\dpp)^2 +\dpm^2)] +  
[(\dtp)^2 +(\dtm)^2 ] +  2\Dlt( \dtm \dpp - \dtp\dpm) \} .
\eeq
In terms of the 
standard spin and  charge
fields defined by $\theta_{c/s}= \frac{\theta_+ \pm \theta_-}{\sqrt
2}$ and $\phi_{c/s} = \frac{\phi_+ \pm \phi_-}{\sqrt
2}$, the Hamiltonian can be rewritten as 
\beq
H = \frac{n\pi}{4}\int dx \{ (1+\Dlt^2)[ (\dpc)^2 +\dps^2)]+ 
[(\dtc)^2 +(\dts)^2 ] + 2\Dlt( \dtc \dps - \dts\dpc) \}.
\eeq
Note that 
unlike the usual models of LLs,  here, the Hamiltonian is not
separable in terms of the spin and charge variables - the $\Dlt$ term
couples them, just as it couples the $\pm$ fields.  This is yet
another motivation to study transport in this model. Until now, the 
Kane-Fisher results\cite{KF}
have only been obtained in models where
spin and charge are explicitly separable. Note also that 
we may work either with  spin and charge fields ($\al=c/s$ in Eq.(\ref{Hcs}))
or  with the $+$ and $-$ fields ($\al=\pm$ in Eq.(\ref{Hcs})). Both
representations are equivalent. 

The Hamiltonian can be brought to a form where it looks almost
non-interacting by defining new fields 
\bea
\tilde \theta_{\pm}  =\partial_x \theta_{\pm} \mp \Dlt \phi_{\mp}
~~~~ {\rm and} ~~~ \tilde \phi_{\pm}= \phi_\pm ,\\
{\rm or} ~~~~\tilde \theta_{c/s}  =\partial_x \theta_{c/s} \pm \Dlt \phi_{s/c}
~~~~ {\rm and} ~~~ \tilde \phi_{c/s}= \phi_{c/s} .
\label{tilde}
\eea
The new fields $\tilde\theta_\pm$ and $\tilde\phi_\pm$ 
($\tilde\theta_{c/s}$ and $\tilde\phi_{c/s}$) satisfy  
the same commutation relations as the non-tilde fields, 
 - $[\tilde \theta_{\al}(x),\tilde \phi_{\al}(y)]
=i\Theta(x-y)=[\theta_{\al}(x),\phi_{\al}(y)]$.
In terms of these fields, 
the Hamiltonian apparently looks spin-charge (or equivalently $+/-$)
separable and is given by 
\beq
H = \frac{n\pi}{4}\sum_{\al=\pm~or~c/s} \int dx \{ 
(\partial_x\tilde \phi_\al)^2 + (\partial_x\tilde \theta_\al)^2 \}.
\label{Hcs}
\eeq
Either one of the fields  can be treated as the
coordinate and the gradient of the other field can be treated 
as the momentum. 

We now study the effects of impurity
scattering in this bosonised model. We will work in the limits of a weak
barrier and a weak link and then extrapolate to intermediate values of
the coupling strength  of the impurity.

Let us first consider scattering from a small barrier. 
A potential scatterer or impurity in the model can be introduced as 
\beq
\delta H =  V ( \psi_+^{\dagger}(0)\psi_+(0)
+\psi_-^{\dagger}(0)\psi_-(0) )
\eeq
where the first and second terms backscatter $+$ and $-$ particles 
respectively. In momentum
space, this term translates into $2k_F$ (and higher momentum
transfer) scatterings between the left and right Fermi points for both
the $+$ and $-$ particles. The
lowest order $2k_F$ scattering term is given by
\beq
\delta H_1 = V_e \sum_{\sigma=\pm} \psi_{\sigma,L}^{\dagger}\psi_{\sigma,R}
+h.c.
\eeq
where  either a $+$
or $-$ particle is back scattered. Here, both charge and spin degrees
of freedom are involved. However, we can also consider higher
order scatterings ($e.g.$, $4k_F$ scattering) where both  $+$ and $-$
particles are scattered
such as 
\beq
\delta H_2 = (V_c~
\psi_{+,L}^{\dagger}\psi_{+,R}\psi_{-,L}^{\dagger}\psi_{-,R} + 
V_s~\psi_{+,L}^{\dagger}\psi_{+,R}\psi_{-,R}^{\dagger}\psi_{-,L}) + h.c. 
\eeq
In the  first(second) term in $\delta H_2$
both kinds of particles are incident from same(different) direction. Thus
the spin(charge) momentum is unaffected and hence, as we shall see, 
the corresponding 
bosonised operator depends only on charge(spin)-fields. 

We will study the effects of both the $2k_F$ and $4k_F$ scattering using
bosonisation and renormalisation group.   
The fermions may be bosonised using 
$$\psi_{R/L}(x)=\lim_{\al\to0}{1\over (2\pi\alpha)^{1/2}}
 e^{\pm i\sqrt{4\pi}\phi_{R/L}}
=\lim_{\al\to0}{1\over (2\pi\alpha)^{1/2}}
 e^{i\sqrt\pi(\pm\phi -\theta)},$$ where $\al$ is a regularisation
parameter, which is finally set to zero. 
(We ignore Klein factors, because we are
using the notation of Shankar\cite{SHANKBOS} which does not require
the use of Klein factors. For an explicit comparison of different
bosonisation schemes, see Ref.\cite{REFERM}).  
Thus, the quadratic and
the quartic fermion operators are bosonised as 
\bea
O_\pm&=&\psi_{\pm,L}^{\dagger}\psi_{\pm,R} ={1\over 2\pi\alpha}
e^{i\sqrt{4\pi}\phi_\pm}
={1\over 2\pi\alpha}e^{i\sqrt{2\pi}(\phi_c \pm\phi_s)}
\label{opm}\\
 {\rm and} \quad
& &O_{c/s}=\psi_{+,L}^{\dagger}\psi_{+,R}\psi_{-,L/R}^{\dagger}\psi_{-,R/L}
={1\over (2\pi\alpha)^2}e^{i\sqrt{8\pi}\phi_{c/s}}.
\eea
Note that as we had expected, the two-fermion operators depend both on
charge and spin, whereas the four-fermion operators involve either the
charge or the spin. 

The renormalisation group flows of these operators to 
leading order can be found by computing their  dimensions. Since the
operators depend on the $\phi$ fields, their dimensions are
conveniently computed using $\phi$ as the coordinate field. For the
bilinear operators, the dimensions are computed as
\beq
<O_\pm^\dagger(\tau,0)O_\pm(0,0)>= \frac{1}{(2\pi\alpha)^2}
(\frac{\alpha^2}{\alpha^2+\tau^2})  
\eeq
whereas for the operators that are quartic in the Fermi fields, we get
\beq
<O_{c/s}^\dagger(\tau,0)O_{c/s}(0,0)>= \frac{1}{(2\pi\alpha)^4}
( \frac{\alpha^2}{\alpha^2+\tau^2})^2.
\eeq
Clearly, the correlation functions diverge as $\tau^{-2}$ for
$O_{\pm}$ and $\tau^{-4}$ for $O_{c/s}$ as 
$\tau \longrightarrow 0$. 
Thus, $O_{\pm}$ are marginal and $O_{c/s}$ are irrelevant, just as
they are 
in the noninteracting case. Generically, once the barrier 
becomes stronger, the above  leading order 
analysis of the RG flow is no
longer valid, and it is
necessary to study the RG flow to higher orders in $V$. 
Since only one operator is marginal and the rest are irrelevant, 
one needs to ask whether the
marginal operator
is marginally relevant or marginally irrelevant when computed to 
higher orders in
$V_e$.  But
here, it is easy to check that the operator $\cos\phi_{\pm}$ can never
contribute to itself at any order in perturbation theory, because
$\cos^n (\sqrt{4\pi}\phi_{\pm})$ does not contain a factor 
$\cos(\sqrt{4\pi}\phi_{\pm})$ for any
$n>1$. Hence, the operator in Eq.(\ref{opm}) or the $2k_F$
backscattering remains marginal to all
orders in $V_e$ and all higher order scattering processes ($4k_F$, 
$6k_F$, ...$etc$.) are irrelevant.

One may still wonder whether the RG flow equation for the $2k_F$ 
backscattering can receive
non-perturbative contributions.  But in this model,
since the '$+$ particles' interact only with '$-$ particles' and 
vice versa, the back scattering of any one kind of particle is just not
affected by the interaction. Another way of seeing this is to note that 
this interacting system
differs from the noninteracting one, by a pseudo unitary transformation 
which does not affect the diagonal density 
operator $ \psi_\sig^\dagger\psi_\sig$.
Hence, even non-perturbative corrections to the lowest order 
flow equations cannot occur. However, we can also see this explicitly if,
following Kane and Fisher\cite{KF}, we study  the model perturbatively
from the other
limit  - $ie.$, the case where there is an
infinite barrier with no transmission across the barrier. 
Then, the fact that the barrier is not infinite,
is modelled by a hopping term across the site. We then use
perturbative RG to determine whether the hopping term is relevant 
leading to a `healing' of the wire, is irrelevant confirming
the stability of the infinite barrier fixed point or is marginal.

In first quantised language, for an infinite barrier, 
the wave function $\psi_\sig$ has to vanish at
the impurity site $x=0$. This allows only odd parity states ($ \hf [
\psi_\sig(x)-\psi_\sig(-x)]$). 
In terms of the bosonised fields, the
condition of infinite barrier at the origin is imposed in the
following way.
We bosonise the $0<x<L/2$  section by introducing the  
bosonic fields $\phi_{\al>}$ and
$\theta_{\al>}$, and  the $0>x>-L/2$ section by  the fields $\phi_{\al<} $ and
$\theta_{\al<}$. (The ring boundary condition is satisfied by
demanding continuity of charge and current at $-L/2 \equiv +L/2$.) 
The Hamiltonian for the infinite barrier case is then given by
$ H=\sum_i H_i$, with $i=>,< $ where,
\bea
\nonumber
H_i & = & \frac{n\pi}{4}\int dx \{ (1+\Dlt^2)[ (\dppi)^2 +\dpmi^2)] +  
[(\dtpi)^2 +(\dtmi)^2 ]  \\ 
 &+&2\Dlt( \dtmi \sum_j\partial_x\phi_{+j}  - 
\dtpi\sum_j\partial_x\phi_{-j}) \} .
\label{Hi}
\eea
with the boundary conditions $\phi_<(0) = \phi_>(0) = \frac{\pi}{2}$
\cite{SCHULZ}. 
Although, we have effectively decoupled the wire by allowing for two
different kinds of bosons $<$ and $>$ to the left and right of the
origin respectively, note that 
the $\sum_j\partial_x\phi_{\sig j}$ terms in the Hamiltonian, 
ensures that the '$+$' particles in any one section ($>$ or $<$),
interacts with all '$-$' particles( $<$ and $>$) in the ring\cite{US}. This 
would  certainly not have been true if we had two disconnected wires(rings).
Since the  interaction energy is not extensive, the original ring 
can never be reproduced starting from 
two completely disconnected rings. The above Hamiltonian, in fact,
models a single ring with no current across the impurity. 
(Here we wrote the Hamiltonian in $\al=\pm$ representation, but
equivalently, we could have written it in terms 
of the spin and charge fields.) A redefinition of both the $<$ and $>$ 
fields analogous to the field redefinitions in Eq.(\ref{tilde}), 
brings it to the apparently  noninteracting form given by 
\beq
H= \frac{n\pi}{4}\sum_{\sig,i} \int dx \{ 
(\partial_x\tilde \phi_{\sig i})^2 + (\partial_x\tilde \theta_{\sig i})^2\} .
\label{Hal}
\eeq 
with $\sig=\pm ~or~c,s$ and $i=<,>$. 

We now introduce a hopping term across $x=0$, as $\delta H= t\sum_\sig
\bar O_{\sig }$, where,
\bea
\nonumber
\bar O_{\sig }&=& \psi_{\sig>}^\dagger(0)\psi_{\sig<}(0) + h.c
={1\over (2\pi\alpha)} e^{i\sqrt{\pi}(\theta_{\sig>}(0) 
-\theta_{\sig<}(0))} + h.c \\
&=&{1\over (2\pi\alpha)} e^{i\sqrt{\pi}(\tilde\theta_{\sig>}(0) 
-\tilde \theta_{\sig<}(0))} + h.c 
\eea
 and $t$ is an overlap matrix element.
Note that since the hopping operator depends only  the $\theta_\sig(0)$
(since $\phi_\sig(0)=0$),  we
may compute its dimension using the Hamiltonian in Eq.(\ref{Hal})
by treating the $\tilde\theta$ fields as the coordinate fields.
The dimension of the hopping operators are easily computed as 
\beq
<\bar O_{\sig }^\dagger(\tau,0)\bar O_{\sig }(0,0)>=
{1\over (2\pi\alpha)^2}
( \frac{\alpha^2}{\alpha^2+\tau^2}),     
\eeq
which says that the operator is marginal as expected, and hence the
single particle backscattering process ($2k_F$) does not depend on the
interaction.  As for the non-interacting case, it depends on the 
strength of the scattering potential and the
energy of the incident particles. In 
this limit too, one can show explicitly that the higher order
processes are irrelevant.

Hence, the result that we obtain for transmission when the wire has a
non-magnetic impurity at the origin introduced either through a weak
$\delta$- function barrier or in the opposite limit, through weak
hopping, is that the strength of the non-magnetic impurity
$V$ is unaffected by the interaction.  The system thus behaves
like a non interacting electron system.  For such impurities,  
depending on the values of the energy and $V$, a part of the incident
electrons gets transmitted  and a part gets reflected. Neither perfect 
transmission nor perfect reflection is  observed.

As an aside, note that the dimension of the same hopping operator can
also be computed in a  system of two completely disconnected rings. 
The Hamiltonian for such a system is the same as Eq. (\ref{Hal}), 
except that the
$\sum_j\partial_x\phi_{\sig j} $ is replaced by $\partial_x\phi_{\sig
i}$. In this model, the dimensions are given by 
\beq
<\bar O_{\sig}^\dagger(\tau,0)\bar O_{\sig }(0,0)> =
{1\over (2\pi\alpha)^2}
( \frac{\alpha^2}{\alpha^2+\tau^2})^{1+\Dlt^2},     
\eeq
which implies that the  operator is  irrelevant. So the model with two
disconnected rings does {\it not} get healed;  rather  it flows to the
insulating fixed point where the two rings  are disconnected.
 
	What about impurities which can flip the 
spin of the electron at the impurity site?  Such   
magnetic impurities can be introduced through spin dependent 
potentials. For a weak barrier, we may introduce,
\beq
\delta H =  \tilde V ( \psi_+^{\dagger}(0)\psi_-(0) + h.c).
\eeq	
At the lowest order, this  will generate two different $2k_F$ backscattering 
processes, $\psi_{+,L}^{\dagger}\psi_{-,R}$ and $\psi_{+,R}^{\dagger}
\psi_{-,L}$, both of which can be written in the bosonised form, as,
\beq
\OO_\pm =\psi_{\pm,L}^{\dagger}\psi_{\mp,R}=e^{i\sqrt\pi(
\phi_\pm +\phi_\mp + \theta_\pm - \theta_\mp) }
\eeq
where $\OO_\pm$ is just the notation for the operators 
$\psi_{+,L}^{\dagger}\psi_{-,R}$ and $\psi_{-,L}^{\dagger}
\psi_{+,R}$ respectively. Since for computing dimensions, it is 
more convenient to use the apparently non-interacting form of 
the Hamiltonian given  in Eq.(\ref{Hcs}), these operators may be
rewritten in terms of the tilde fields as
\beq
\OO_\pm =  
e^{i\sqrt\pi[(1\pm \Dlt )\tilde\phi_\pm +(1\pm \Dlt )\tilde \phi_\mp
 -\tilde\theta_\pm +\tilde\theta_\mp) }
\eeq
using Eq.(\ref{tilde})
The lowest order contribution to the RG equation for $V$ is now simply
obtained by computing the dimension of the operator. 
The correlation function is now given by 
\beq
<\tilde O_\pm^\dagger(\tau,0)\tilde O_\pm(0,0)>=
{1\over (2\pi\alpha)^2}
( \frac{\alpha^2}{\alpha^2+\tau^2})^{\hf[1+(1\pm\Dlt)^2]},     
\eeq
	and diverges as $\tau^{-2-\Dlt^2\mp 2\Dlt }$  as 
$\tau \longrightarrow 0$. So the operator 
$\OO_+ = \psi_{+,L}^{\dagger}\psi_{-,R}$ 
is relevant for $\Dlt < 0$, whereas the operator  
$\OO_-=\psi_{-,L}^{\dagger}\psi_{+,R}$ is relevant for  
 $\Dlt > 0$. Thus one of the two backscattering 
spin flip operators  is always 
relevant for either sign of effective interaction strength. 
Hence, the magnetic impurity cuts the wire at the site 
of the impurity. Since  the lowest order $2k_F$ backscattering
is relevant, we need not look for higher order processes. An 
explicit check, however, shows that all such operators are irrelevant.
	Once the impurity strength grows under
renormalisation, the weak barrier analysis is no longer valid. 
As was done in the case with non-magnetic impurities, we need to study
the strong barrier limit, where we start from the insulating limit and
then study weak hopping. We take the Hamiltonian of Eq. 
(\ref{Hi}) in the $\theta$ representation, (or Eq.(\ref{Hal}) in the
$\tilde\theta$ representation)
which is effectively
cut at the site of impurity as the unperturbed hamiltonian.
The hopping of electrons 
across the the weak link, accompanied by a spin flip is modelled by  
$$
\delta H =  \tilde t( \psi_{+>}^{\dagger}(0)\psi_{-<}(0)
+ h.c. ),
$$
 which can be bosonised as follows,
\bea
\delta H= \frac{\tilde t}{2\pi\al}
e^{i\sqrt{\pi}(\theta_{+>}-\theta_{-<})}+h.c
= \frac{2\tilde t}{2\pi\al}cos[\sqrt{\pi}(\tilde\theta_{+>}-\tilde\theta_{-<}-
\Dlt\sum_{\sig,i}\tilde\phi_{\sig,i})]
\eea
where the apparently non-interacting tilde fields have been used in
the second line. The scaling dimension can easily be read off as 
$2+4\Dlt^2$. So  hopping remains irrelevant for all
values (both +ve and -ve) of the interaction parameter.
Here again, we can explicitly check that all 
higher order processes continue to be more irrelevant. 
The wire which is cut at the weak barrier limit, 
does not get 'healed' for any strenght of the barrier. 
This  interacting  electron system, thus, flows
towards an insulating fixed point in the presence of magnetic 
impurity. 

Let us contrast these results  with the results obtained  in a purely
TR invariant model of a Luttinger liquid, -
$e.g.$, the model studied by Kane and Fisher\cite{KF}.  They found that
for spinless fermions, repulsive interactions imply that any barrier
cuts the wire and attractive interactions imply that 
any barrier is rendered
invisible in the renormalisation group sense. However, once both spins and
charges are introduced with arbitrary
values of $g_{\rho}$ and $g_{\sigma}$ denoting the strengths of the
interactions of the charge and spin sectors, various phases
do exist. Besides the purely conducting and the purely insulating phases,
mixed phases where there exists a non-trivial fixed point depending on
barrier height separating the two phases also exist. 
Kane and Fisher also showed  how to access the non-trivial fixed points 
by studying the renormalisation group equations to second order in the
impurity strengths around the weak and the strong barriers.

This model cannot be fitted into the general framework of the
Kane-Fisher models of Luttinger liquids, because it explicitly
violates TR symmetry. However, at a qualitative level, the results are
similar to the Kane-Fisher results\cite{KF} for repulsive
interactions.  At least for spin-flip impurities, one finds that any
barrier, however small, cuts the wire. For non-magnetic or
spin-conserving impurities, however, we found that this model is
similar to a non-interacting model, and allows both transmission and
reflection.Note, however that we are not assuming phase coherence 
through the whole ring as exists for mesoscopic systems. 

In this model, the strengths of the interactions of both the spin and
the charge degrees of freedom are completely fixed by the gauge
interaction. We have no freedom in changing the relevant strengths of
the interactions, or in fact, explicitly changing the strength  of the
TR violation. However, in a more general model of transport which
allows for arbitrary TR symmetry violation\cite{USNEXT}, 
we can try to look for other phases such as the purely conducting
phase or mixed phases separated by non-trivial fixed points. 
It should also be possible in
such models to gradually change the strength of the TR violation
and show that in the limit where the TR violation  goes to zero,
the Kane-Fisher results are recovered. Another useful
generalisation\cite{USNEXT} 
is to study transport in many-chain(band) models which can be modelled
for some purposes by a
multi-band Schulz-Shastry model\cite{US,SS2}. Here, again, it may be
possible to have more phases and perhaps non-trivial fixed points
separating the phases.

In conclusion, we have studied transport in a a simple exactly
solvable model of interacting fermions 
and have obtained the anomalous dimensions of several back-scattering
operators in the weak barrier and weak link limits. Since at low
temperatures and low driving forces, the conductance has a power law
behaviour in terms of the anomalous dimensions, we can directly 
obtain results for transmission
and conductance.
With several new experiments on one dimensional wires (semi-conductor wires
in the single channel limit, single wall carbon nano-tubes, {\it etc.}),
it becomes essential to  include interactions.  The usual 
Landauer-Buttiker formalism will not suffice. Hence, it is important to
study transport in models with  interactions. Besides 
the results that one gets in the standard Luttinger model, 
it will definitely
be extremely useful to have other models of
interacting fermions, where
many of these ideas of transport through LLs
in various geometries can be tested. Since any external voltage
applied to the wires breaks TR invariance, it is also of interest
to study transport in models where TR symmetry  is explicitly violated. 
In this context, we feel 
that the exactly solvable models proposed by Schulz and Shastry 
could serve as a useful tool.



\begin{thebibliography}{99}

\bibitem{1DWIRE} T.J. Thornton {\it et al.}, \prl  {\bf56},
1198 (1986); A. Yacoby {\it et al.}, \prl  {\bf 77}, 135 (1996);
K. J. Thomas {\it et al.}, \prl  {\bf 77}, 4612 (1996).

\bibitem{CTUBES} S. Ijima, Nature {\bf 354}, 56 (1991); A. Thess {\it
et al}, Science {\bf 273}, 483 (1996); S. J. Tans {\it et al.}, Nature
{\bf 386}, 474(1997) .

\bibitem{DEKKER}T. W. Ebesen {\it et al.}, Nature {\bf 382}, 54 (1996); 
D. H. Cobden {\it et al.}, \prl  {\bf 81}, 681 (1998).

\bibitem{HA}
For a recent introduction to one-dimensional systems, see Zachary
N. C. Ha, `Quantum Many-Body systems in One Dimension', World
Scientific, Singapore, 1998.


\bibitem{KF}
C. L. Kane and M. P. A. Fisher, \prb{\bf 46}, 15233 (1992).

\bibitem{GR}
D. L. Maslov and M. Stone, \prb{\bf
52}, R5539 (1995); 
I. Safi and H. J. Schulz, \prb{\bf 52} R17040 (1995);
A. Furusaki and N. Nagoasa, \prb{\bf 54}, R5239 (1996).

\bibitem{LUDWIG}
P. Fendley, A. W. Ludwig and H. Saleur, \prb{\bf 52}, 8934 (1995).


\bibitem{EG}
R. Egger and H. Grabert, \prb{\bf 58}, 10761 (1998).



\bibitem{KE}
A. Komnik and R. Egger, \prl  {\bf 80}, 2881 (1998)

\bibitem{DR}
P. Durganandini and S. Rao, \prb{\bf 59}, 13122 (1999); {\it ibid },
 MRI-PHY/P991032.  

\bibitem{SS}
H. J. Schulz and B. S. Shastry, \prl  {\bf 80}, 1924 (1998).

\bibitem{US}
R. K. Ghosh, P. K. Mohanty and S. Rao,
Jnl.  of Phys. {\bf A32}, 4343 (1999). 

\bibitem{SS2}
H. J. Schulz and B. S. Shastry, \prl  {\bf 82}, 2410 (1999).

\bibitem{BOS}
For a review, see J. Voit, Rep. Prog. Phys. {\bf 57}, 977 (1995);
other reviews include
H.J.Schulz, Int.J. of Mod. Phys. {\bf B5}, 57 (1991) and  
M. Stone, "Bosonization", World Scientific, 1994. 

\bibitem{SHANKBOS}
R.Shankar, `Bosonisation: How to make it work for you in condensed
matter' , in Low-Dimensional Quantum Field Theories for Condensed
Matter Physics', World Scientific, 1992.


\bibitem{REFERM}
For a detailed comparison of the various bosonisation schemes, see 
J. von Delft and H. Schoeller, Annalen Phys. {\bf 7}, 225 (1998).

\bibitem{SCHULZ} H. J. Schulz, 'Fermi Liquids and Non-Fermi Liquids',
in 'Mesoscopic Quantum Physics', Les Houches 1994,
North-Holland.

\bibitem{USNEXT} 
P. K. Mohanty and S. Rao, work in progress.

\end{thebibliography}
\end{document}